\documentclass[bibyear]{aa} 

\usepackage{txfonts}
\usepackage{rotating}
\usepackage{amssymb}
\usepackage{color}
\usepackage{fancyhdr}
\usepackage{textcomp}
\usepackage{longtable}
\usepackage{tabularx}
\usepackage{graphicx}
\usepackage{natbib}
\usepackage{amsmath}
\usepackage{mathtools}
\usepackage[ampersand]{easylist}
\usepackage{mathrsfs}

\usepackage[usenames,dvipsnames,svgnames,table]{xcolor}

\def \mlk {$\rm L_K-M_{WL}$ \,}

\begin{document} 

   \title{The XXL Survey X: K-band luminosity - weak-lensing mass relation for groups and clusters of galaxies}
 \titlerunning{K-band luminosity - weak lensing mass relation}  
   \subtitle{}

   \author{F. Ziparo
          \inst{1}
          \and
                        G. P. Smith 
                        \inst{1}
            \and
                    S. L. Mulroy
                        \inst{1}
                        \and    
                        M. Lieu
            \inst{1}
                        \and    
                        J. P. Willis
                        \inst{2}
                        \and
                        P. Hudelot
                        \inst{3}
                        \and
                        S.~L.~McGee
                        \inst{1}
                        \and
                        S. Fotopoulou
                        \inst{4}
            \and
            C. Lidman
            \inst{5}
                        \and
                        S. Lavoie
                        \inst{2}
                        \and
                        M.~Pierre
                        \inst{6}
            \and
                        C.~Adami
                        \inst{7}
            \and        
                        L.~Chiappetti
                        \inst{8}
            \and
                        N.~Clerc
                        \inst{9}
                        \and                                    
                        P.~Giles
                        \inst{10}
                        \and
                        B.~Maughan
                        \inst{10}
            \and
                        F.~Pacaud
                        \inst{11}
                        \and    
                        T. Sadibekova
                        \inst{6}                        
          }

   \institute{School of Physics and Astronomy, University of Birmingham, Edgbaston, Birmingham B15 2TT, UK \\
              \email{fziparo@star.sr.bham.ac.uk}
                          \and
                          Department of Physics and Astronomy, University of Victoria, 3800 Finnerty Road, Victoria, BC, V8P 1A1, Canada
                          \and
                          Institut d'Astrophysique de Paris, CNRS UMR 7095 and UPMC, 98bis, bd Arago, F-75014 Paris, France 
                          \and
                          Department of Astronomy, University of Geneva, ch. d'\'Ecogia 16 CH-1290 Versoix
                          \and
                          Australian Astronomical Observatory, PO Box 915, North Ryde, NSW 1670, Australia
                          \and
                          Service d’Astrophysique AIM, CEA Saclay, F-91191 Gif sur Yvette
                          \and
                          Aix Marseille Universit\'e, CNRS, LAM (Laboratoire d'Astrophysique de Marseille) UMR 7326, 13388, Marseille, France
                          \and
                          INAF, IASF Milano, via Bassini 15, I-20133 Milano, Italy
                          \and
                          Max-Planck-Institute for Extraterrestrial Physics, Giessenbachstrasse 1, 85748, Garching, Germany
                          \and
                      H. H. Wills Physics Laboratory, University of Bristol, Tyndall Avenue, Bristol, BS8 1TL, England       
                          \and
                          Argelander-Institut für Astronomie, University of Bonn, Auf dem Hügel 71, 53121 Bonn, Germany
             }

   \date{Received -- ; accepted --}

 
  \abstract
   {
   Galaxy clusters and groups are important cosmological probes and giant cosmic laboratories for studying galaxy evolution. Much effort has been devoted to understanding how and when baryonic matter cools at the centre of potential wells. However, a clear picture of the efficiency with which baryons are converted into stars is still missing. 
We present the K-band luminosity--halo mass relation, $\rm L_{K,500}-M_{500,WL}$, for a subsample of 20 of the 100 brightest clusters in the XXL Survey observed with WIRCam at the Canada-France-Hawaii Telescope (CFHT). For the first time, we have measured this relation via weak-lensing analysis down to $\rm M_{500,WL} =3.5 \times 10^{13}\,M_\odot$.
This allows us to investigate whether the slope of the $\rm L_K-M$ relation is different for groups and clusters, as seen in other works.
The clusters in our sample span a wide range in mass, $\rm M_{500,WL} =0.35-12.10 \times 10^{14}\,M_\odot$, at $0<z<0.6$. The K-band luminosity scales as $\rm \log_{10}(L_{K,500}/10^{12}L_\odot) \propto \beta \log_{10}(M_{500,WL}/10^{14}M_\odot)$ with $\beta = 0.85^{+0.35}_{-0.27}$ and an intrinsic scatter of $\rm \sigma_{lnL_K|M} =0.37^{+0.19}_{-0.17}$. Combining our sample with some clusters in the Local Cluster Substructure Survey (LoCuSS) present in the literature, we obtain a slope of $1.05^{+0.16}_{-0.14}$ and an intrinsic scatter of $0.14^{+0.09}_{-0.07}$. 
The flattening in the $\rm L_K-M$ seen in previous works is not seen here and might be a result of a bias in the mass measurement due to assumptions on the dynamical state of the systems.
We also study the richness-mass relation and find that group-sized halos have more galaxies per unit halo mass than massive clusters. However, the brightest cluster galaxy (BCG) in low-mass systems contributes a greater fraction to the total cluster light than BCGs do in massive clusters; the luminosity gap between the two brightest galaxies is more prominent for group-sized halos. 
This result is a natural outcome of the hierarchical growth of structures, where massive galaxies form and gain mass within low-mass groups and are ultimately accreted into more massive clusters to become either part of the BCG or one of the brighter galaxies.
}



\maketitle
%


\section[]{Introduction}
\label{sec:intro}

Galaxies grow through predominantly dissipative processes within dark matter halos \cite[e.g.][]{White_Rees1978, Davis1985, Springel2006}; according to the standard paradigm of cosmological structure formation, these galaxies assemble first in groups and then in clusters and are the largest objects to form in the process. Galaxy clusters are important both as cosmological probes and as laboratories for galaxy evolution studies \cite[see e.g.][for a recent review]{Kravtsov2012}.

It is still not understood how baryonic matter cools and fragments at the centre of gravitational potential wells to trigger star formation and increase galaxy stellar mass. 
Much effort has been devoted to the study of stellar versus halo mass to constrain the star formation efficiency of galaxy groups and clusters. 
In particular, many works present in the literature find that stellar and halo mass are closely correlated with a slope shallower than unity \cite[e.g.][]{Giodini2009,Patel2015,Andreon2010,Balogh2014,Kravtsov2014,VanDerBurg2014}. This implies that group-sized halos are more efficient at forming stars than their more massive counterparts \cite[see also][]{Gonzalez2013}. 

A similar conclusion has been reached by earlier authors using near-infrared luminosity and in particular the K-band luminosity $\rm L_K$ as a tracer of stellar mass \cite[e.g.][]{Lin2003,Lin2004,Ramella2004,Muzzin2007}. There are many advantages of using $\rm L_K$ instead of stellar mass: it is relatively inexpensive compared to multiwavelength observations; it is not  affected very much by extinction or recent star formation \citep{Cowie1994}; and k-corrections are small and almost independent of galaxy type \cite[e.g.][]{Poggianti1997}.  
The Two Micron All-Sky Survey \cite[2MASS;][]{Jarrett2000} offers a complete infrared map of the sky, which enables the study of large sample of galaxies, groups, and clusters \cite[e.g.][]{Lin2003,Lin2004,Kochanek2001,Ramella2004}. The K-band luminosity--mass ($\rm L_K-M$) relation for local systems revealed a slope of 0.6 with system masses measured via dynamical analysis \citep{Ramella2004} or inferred from X-ray scaling relations \citep{Lin2004}. \cite{Muzzin2007} have explored the $\rm L_K-M$ for massive clusters in a wide redshift range ($0.17<z<0.54$), finding consistent results with other work in the local Universe. They conclude that there is little evolution in the $\rm L_K-M$ relation with redshift and cluster mass.

All the aforementioned studies estimate cluster masses via X-ray scaling relations or dynamical analysis. This means that they rely on assumptions on the dynamical state of the systems. \cite{Mulroy2014} investigated, for the first time, the $\rm M-L_K$ relation using weak-lensing masses. Their sample from the Local Cluster Substructure Survey (LoCuSS\footnote{http://www.sr.bham.ac.uk/locuss}) comprised clusters with masses of $\rm M_{500,WL}> 2 \times 10^{14}~M_\odot$. These authors find a slope of unity and an intrinsic scatter of $\rm \sigma_{lnM|L_K}=0.1$. 
The difference in slope between the Mulroy et al. lensing-based study of massive clusters and the literature indicate that the slope of the  $\rm M-L_K$ relation is a function of halo mass and/or that earlier studies were affected by systematic uncertainties in mass measurements.

Weak-lensing analysis is difficult to perform for low-mass systems, mainly owing to the weakness of the signal. For this reason there are no studies on $\rm L_K-M$ with masses measured via weak-lensing  ($\rm M_{WL}$) down to group-sized halos. In this work we present the first $\rm L_K-M_{WL}$ relation down to $\rm \sim 10^{13}~M_\odot$ for 20 of the 100 brightest clusters in the XXL\footnote{http://irfu.cea.fr/xxl} Survey. This pilot study aims to explore the slope of the \mlk relation for a wide range of masses to shed light on their star formation efficiency without relying on any hydrostatic equilibrium assumptions. This allows us to investigate, for the first time, whether the change in slope of the \mlk relation is a function of halo mass or whether it is due to a bias in the mass measurements in previous works.

The paper is organised as follows: in \textsection~\ref{sec:data} we describe our data set and sample; in \textsection~\ref{sec:analysis} we introduce our method for measuring the K-band luminosity and for investigating the \mlk and the richness-mass relations in \textsection~\ref{sec:results}; we discuss our results and compare them with other works present in the literature in \textsection~\ref{sec:discussion}; and we draw our conclusions in \textsection~\ref{sec:conclusions}.
Throughout our analysis we adopt the Vega magnitude system and the WMAP9 \citep{Hinshaw2013} cosmology of $\rm H_0 = 70\, h_{70}\, km\, s^{-1}\, Mpc^{-1}$, $\rm \Omega_M = 0.28$, and $\rm \Omega_\Lambda = 0.72$.


\section[]{Sample and data}
\label{sec:data}

The XXL Survey, described in detail by Pierre et al. (submitted, hereafter Paper I), is a $\rm 50~deg^2$ XMM-Newton survey with a sensitivity of $\rm \sim 5\times 10^{-15} erg\, s^{-1}\, cm^{-2}$ in the [0.5-2]~keV band for point-sources. This survey is an extension of the $\rm 11~deg^2$ XMM-LSS survey \citep{Pierre2004} and consists of two regions of $\rm 25~deg^2$ each, XXL-North and XXL-South. 
The main aim of XXL is to provide a well-defined galaxy cluster sample for studies of precision cosmology, galaxy evolution, and active galactic nuclei.

Within the XXL Survey, the bright XXL 100 cluster sample (XXL-100-GC\footnote{XXL-100-GC data is available as electronic format via the XXL Master Catalogue at \url{http://cosmosdb.iasf-milano.inaf.it/XXL}} and via the XMM XXL DataBase at \url{http://xmm-lss.in2p3.fr}.) is defined as a flux-limited sample based on the 100 brightest clusters (more details available in Pacaud et al. (submitted, hereafter Paper II). Some of these clusters were previously studied as part of the XMM-LSS and XMM-BCS surveys \citep{Clerc2014,Suhada2012} and  span a wide redshift range ($0.05\lesssim z \lesssim 1.07$). 
All systems within the XXL-100-GC sample are characterised as either C1 or C2 classification. All C1 objects have high extension and detection likelihood with a low probability of contamination by spurious detection or point sources, while C2 objects are much less pure, with only half of the sources corresponding to real clusters. All of the clusters used in this work are C1 systems, although we stress that we did not base our selection on whether clusters are ranked as C1 or C2.

As described in \cite{PaperI} and \cite{PaperII}, an extensive follow-up programme has been carried out to obtain spectroscopic redshifts for all XXL galaxy clusters. We assume that a cluster is spectroscopically confirmed if three consistent redshifts lie within 500~kpc from the X-ray centroid or if at least the BCG has a spectroscopic redshift. The spectroscopic redshifts for the systems studied in this work are listed in Table~\ref{table:data}, while the number of spectroscopic members can be found in Table~1 of \cite{PaperII}).

\subsection{Sample}
In this work we use a sample of clusters drawn from the overlap between XXL-100-GC, CFHTLenS \citep{Erben2012,Heymans2012}, and MIRACLES (a wide near-infrared survey covering a large part of the XXL-N field with WIRCam observations, Arnouts et al., in prep.), for which reliable weak-lensing masses are available from Lieu et al. (submitted, hereafter Paper IV). \citet{Lieu2015} selects a sample of 38 galaxy clusters for which redshifts \citep{PaperII}; faint galaxy shape measurements; and X-ray temperatures, $\rm T_X$ (Giles et al. submitted, hereafter Paper III), are available. 
From these 38 clusters we select all the clusters with K-band data, which results in a sample of 20 clusters (see Table~\ref{table:data}) all of which are classified as C1 (thus with a reliable X-ray detection) and have masses in the range $\rm M_{500,WL} =0.35-12.10 \times 10^{14}\,M_\odot$ at $0<z<0.6$.

\subsection{Weak-lensing masses}
\label{sec:wl_masses}
We use weak-lensing masses from \cite{Lieu2015} in the overlapping region with the WIRCam coverage (yielding 20 clusters, Table~\ref{table:data}). All details on weak-lensing analysis and mass measurement can be found in \cite{Lieu2015}. We summarise the main points here.
\cite{Lieu2015} analyses 38 systems drawn from the XXL-100-GC sample, for which the CFHTLenS \citep{Heymans2012} shear catalog is publicly available\footnote{http://www.cfhtlens.org/astronomers/content-suitable-astronomers}.
To obtain weak-lensing masses $\rm M_{500,WL}$, a \cite{NFW1997} profile is fitted to the shear profile of each cluster and integrated to the radius at which the mean density of the halo is 500 times the critical density of the universe at the cluster redshift. In this work we also use masses integrated to the radius of 1~Mpc, $\rm M_{1Mpc,500}$. 

The individual masses measured for the systems span a wide range of masses ($\rm M_{500,WL}\sim 10^{13}-10^{15}$) and temperatures ($\rm 1~KeV \lesssim T_X \lesssim 6~KeV$) and represent the largest sample of groups and clusters with weak-lensing masses for which the mass-temperature relation has ever been studied.

Table~\ref{table:data} lists the main properties of the 20 clusters considered in this work. 

\begin{table*}
\caption{Cluster properties. Column 1 shows the cluster identification number; Col. 2 shows the cluster redshift (from \cite{PaperII}); the weak-lensing masses measured within $\rm r_{500,WL}$ (from \cite{Lieu2015}) and 1~Mpc are shown in Cols. 3 and 4, respectively; Cols. 5 and 6 show the K-band luminosities measured within $\rm r_{500,WL}$ and 1~Mpc from the X-ray centroid of the cluster, respectively; Cols. 7 and 8 are the stellar masses obtained by multiplying the K-band luminosities within $\rm r_{500,WL}$ and 1~Mpc by a fixed mass-to-light ratio of 0.73; Col. 9 is the number of galaxies contributing to the K-band luminosity within $\rm r_{500,WL}$. The position (Right Ascension and Declination) of these objects can be found in Table~1 of \cite{PaperII}. 
}             
\label{table:data}      
\centering                          
\begin{tabular}{cccccccccc}        
\hline
\hline                 
  
XLSSC ID  &  z  &  $\rm M_{500,WL}$   &  $\rm M_{1Mpc,WL}$   & $\rm L_{K,500}$  & $\rm L_{K,1Mpc}$  & $\rm M_{\star,500}$ & $\rm M_{\star,1Mpc}$  &   $\rm N_{500}$   \\   
          &     & ($\rm 10^{14}~M_\odot$)    & ($\rm 10^{14}~M_\odot$) & ($\rm 10^{12}~L_{K,\odot}$)  & ($\rm 10^{12}~L_{K,\odot}$) &   ($\rm 10^{12}~M_\odot$) &   ($\rm 10^{12}~M_\odot$) &      \\
 (1)  &  (2)     &  (3)        & (4)                & (5)          &  (6)  & (7)   & (8)   & (9) \\   
\hline
 027 &  0.295  &  $ 2.1^{+ 2.4} _{- 1.4}$ & $ 2.6^{+ 1.3} _{- 1.7}$ & $ 5.1\pm 1.0$ & $ 5.1 \pm  1.4$ & $ 3.7\pm 0.8$ & $ 3.7 \pm  1.0$ & $ 49\pm15$ \\
 054 &  0.054  &  $ 0.7^{+ 1.1} _{- 0.5}$ & $ 1.1^{+ 0.6} _{- 1.0}$ & $ 2.2\pm 0.5$ & $ 2.6 \pm  1.0$ & $ 1.6\pm 0.4$ & $ 1.9 \pm  0.7$ & $ 39\pm 9$ \\
 055 &  0.232  &  $ 5.2^{+ 4.7} _{- 2.0}$ & $ 4.6^{+ 1.2} _{- 2.1}$ & $ 8.3\pm 1.9$ & $ 7.6 \pm  1.6$ & $ 6.1\pm 1.4$ & $ 5.5 \pm  1.2$ & $ 114\pm26$ \\
 056 &  0.348  &  $ 2.8^{+ 1.7} _{- 1.5}$ & $ 3.2^{+ 1.3} _{- 1.1}$ & $11.1\pm 1.6$ & $11.6 \pm  1.9$ & $ 8.1\pm 1.2$ & $ 8.5 \pm  1.4$ & $ 84\pm15$ \\
 060 &  0.139  &  $ 1.4^{+ 0.9} _{- 1.0}$ & $ 1.9^{+ 1.0} _{- 1.0}$ & $ 7.3\pm 1.1$ & $ 9.6 \pm  1.6$ & $ 5.3\pm 0.8$ & $ 7.0 \pm  1.2$ & $ 103\pm 10$ \\
 061 &  0.259  &  $ 2.4^{+ 0.5} _{- 1.3}$ & $ 2.8^{+ 1.2} _{- 0.4}$ & $ 2.7 \pm 1.3$ & $ 2.8\pm 1.1$ & $ 2.0\pm 0.8$ & $ 1.9 \pm  1.0$ & $ 71\pm18$ \\
 083 &  0.430  &  $ 2.5^{+ 2.2} _{- 1.7}$ & $ 3.1^{+ 1.7} _{- 1.5}$ & $ 8.5\pm 1.5$ & $ 9.5 \pm  1.9$ & $ 6.2\pm 1.1$ & $ 7.0 \pm  1.4$ & $ 87\pm12$ \\
 084 &  0.430  &  $ 2.7^{+ 1.9} _{- 2.0}$ & $ 3.3^{+ 2.0} _{- 1.3}$ & $ 7.5\pm 1.1$ & $ 9.8 \pm  1.4$ & $ 5.5\pm 0.8$ & $ 7.1 \pm  1.0$ & $ 89\pm13$ \\
 087 &  0.141  &  $ 0.3^{+ 0.3} _{- 0.2}$ & $ 0.6^{+ 0.4} _{- 0.4}$ & $ 1.0\pm 0.4$ & $ 1.5 \pm  1.1$ & $ 0.7\pm 0.3$ & $ 1.1 \pm  0.8$ & $ 18\pm 8$ \\
 088 &  0.295  &  $ 1.2^{+ 0.9} _{- 0.9}$ & $ 1.7^{+ 1.1} _{- 1.2}$ & $ 6.9\pm 0.8$ & $ 8.7 \pm  1.4$ & $ 5.0\pm 0.6$ & $ 6.3 \pm  1.0$ & $ 60\pm10$ \\
 091 &  0.186  &  $ 6.2^{+ 2.1} _{- 1.8}$ & $ 4.9^{+ 1.0} _{- 0.9}$ & $18.0\pm 2.2$ & $17.2 \pm  1.8$ & $13.2\pm 1.6$ & $12.5 \pm  1.3$ & $ 233\pm27$ \\
 098 &  0.297  &  $ 1.8^{+ 2.3} _{- 1.5}$ & $ 2.4^{+ 1.6} _{- 1.8}$ & $ 4.0\pm 0.9$ & $ 4.4 \pm  1.3$ & $ 2.9\pm 0.7$ & $ 3.2 \pm  1.0$ & $ 44\pm13$ \\
 103 &  0.233  &  $ 5.4^{+ 2.6} _{- 1.8}$ & $ 4.6^{+ 1.1} _{- 1.2}$ & $ 2.5\pm 1.5$ & $ 2.6 \pm  1.2$ & $ 1.8\pm 1.1$ & $ 1.9 \pm  0.9$ & $ 35\pm25$ \\
 104 &  0.294  &  $ 1.7^{+ 2.6} _{- 0.9}$ & $ 2.2^{+ 0.9} _{- 1.8}$ & $ 4.3\pm 0.8$ & $ 4.6 \pm  1.2$ & $ 3.2\pm 0.6$ & $ 3.4 \pm  0.9$ & $ 50\pm13$ \\
 105 &  0.429  &  $12.1^{+ 3.9} _{- 4.6}$ & $ 8.0^{+ 1.2} _{- 2.5}$ & $ 8.8\pm 2.4$ & $ 8.3 \pm  1.5$ & $ 6.4\pm 1.7$ & $ 6.0 \pm  1.1$ & $ 101\pm30$ \\
 106 &  0.300  &  $ 4.3^{+ 1.8} _{- 2.1}$ & $ 4.1^{+ 1.5} _{- 1.0}$ & $10.9\pm 1.7$ & $11.3 \pm  1.6$ & $8.0\pm 1.2$ & $ 8.2 \pm  1.2$ & $ 129\pm22$ \\
 109 &  0.491  &  $ 4.7^{+ 4.0} _{- 2.8}$ & $ 4.8^{+ 2.1} _{- 2.1}$ & $ 4.4\pm 1.8$ & $ 4.4 \pm  1.8$ &$ 3.2\pm 1.3$ & $ 3.2 \pm  1.3$ & $ 28\pm16$ \\
 110 &  0.445  &  $ 2.9^{+ 3.2} _{- 1.0}$ & $ 3.4^{+ 0.8} _{- 2.0}$ & $ 9.3 \pm  2.2$ & $ 9.6\pm 1.8$ & $ 7.0\pm 1.3$ & $ 6.8 \pm  1.6$ & $ 65\pm13$ \\
 111 &  0.299  &  $ 6.3^{+ 1.8} _{- 1.8}$ & $ 5.2^{+ 1.0} _{- 0.8}$ & $13.2 \pm  1.8$ & $15.0\pm 2.3$ & $11.0\pm 1.7$ & $ 9.7 \pm  1.3$ & $139\pm28$ \\
 112 &  0.139  &  $ 0.8^{+ 0.6} _{- 0.5}$ & $ 1.3^{+ 0.8} _{- 0.5}$ & $ 2.3\pm 0.7$ & $ 1.9 \pm  1.3$ & $ 1.6\pm 0.5$ & $1.4 \pm  0.9$ & $18\pm 10$ \\
 \hline
                   
\end{tabular}
\end{table*}

\subsection{Optical and near-infrared imaging}

This paper uses $ugriz$ imaging data from the Canada-France-Hawaii Telescope Legacy Survey (CFHTLS\footnote{http://www.cfht.hawaii.edu/Science/CFHTLS}) in addition to CFHT WIRCam $K_s$-band data over a subset of the CFHTLS W1 field.  The WIRCam camera consists of four HAWAII2-RG detectors, each containing $2048 \times 2048$ pixels. The four detectors image an area of $20\arcmin \times 20\arcmin$ with a pixel scale of $0.3\arcsec$/pixel (the cross-shaped gap between the four detectors is 45\arcsec\ wide).  A total of 151 WIRCam pointings were observed, each for a total of 1050 seconds with observations acquired as two sequences of 21 spatially dithered exposures of 25 seconds.

The WIRCam data is part of the MIRACLES survey and is reduced subsequently at CFHT and TERAPIX\footnote{http://terapix.iap.fr}. 
This pipeline makes extensive use of the software from the Astromatic webpage\footnote{http://www.astromatic.net} and is similar to the one used to process the CFHT-WIRDS \citep{Bielby2014} and the ESO-UltraVISTA \citep{McCracken2012} surveys.
As a first step, we performed a single exposure detrending using the I'iwi preprocessing pipeline\footnote{http://cfht.hawaii.edu/Instruments/Imaging/WIRCam/IiwiVersion1 Doc.html} with the aim of removing the instrumental imprints from individual images: flagging of saturated, bad and hot pixels, correction for non-linearity, bias removal, guide window masking, and  dome and sky flat fielding. We then computed a first astrometric solution for each exposure and used standard star observations to compute the photometric zero-points.

The initial median-combined stacks constructed using {\tt Swarp} at the instrument pixel scale allowed us to detect faint objects after an initial sky subtraction. The same objects were then masked on individual exposures (taken over time intervals $\rm \Delta t<15$~min and angular separations $\rm \Delta \theta <10^\prime$) to obtain an improved sky background estimate.
The images were then sky-subtracted using these sky frames.

We computed a refined astrometric calibration using {\tt SCAMP} on catalogs detected from the new sky subtracted exposures, reaching an internal astrometric precision of 0.025\arcsec (better than a tenth of a pixel) and an external astrometric precision of 0.18\arcsec (the external astrometric precision is limited by the internal accuracy of the 2MASS-PSC catalog used as reference). Before the final stack, we performed a quality assessment of the individual images to remove those with severe defects. We measured the seeing for each image using {\tt PSFEx}, removing outliers in the image quality before combining the sample and keeping an average seeing of $\sim 0.9\arcsec$. We used {\tt Swarp} to produce the final stacks in four versions, two with different kernels (bilinear and Lanczos3) and two with different grid sizes (native WIRCam and CFHTLS-MegaCam), with a 128px mesh for large-scale background gradient subtraction.

Each exposure was finally delivered with an initial astrometric and photometric calibration in two flavours: the detrended exposures, and the detrended images with the sky removed. 
TERAPIX finally used {\tt QualityFITS} on detrended sky subtracted exposures to produce weight maps, object catalogues, and overviews of individual image qualities (e.g. seeing, depth). The production of the weight maps was possible via the WeightWatcher software \citep{Marmo2008}, while astrometric and photometric calibration was performed using {\tt SCAMP} \citep{Bertin2006} with the Two Micron All Sky Survey (2MASS) taken as the astrometric reference catalogue.

Source extraction and photometry were performed on the $ugrizK_s$ images using {\tt SExtractor v2.5.0} \citep{Bertin1996} in dual image mode with the $K_s$-band as the detection image in each case. Photometry was extracted within fixed circular apertures (3, 4, and 8\arcsec) or in flexible Kron-like elliptical apertures \citep{Kron1980} with a Kron-factor of 2.5 and a minimum radius of 3.5 pixels.  The CFHTLS and $K_s$ image files form an overlapping grid of tiles $\rm 1~deg^2$ in size.  Catalogues generated from each of these tiles were merged into a single master catalogue following the procedures outlines in \cite{Gwyn2012}.
We used a 3\arcsec~ diameter circular aperture to measure the colour of galaxies, while the K-band luminosity was derived from {\tt MAG\_AUTO}, which is a Kron-like magnitude.

To distinguish between stars and galaxies, we used the half-light radius ($r_h$,  defined as the radius within which 50\% of the object flux is enclosed). For all galaxies with $z_{AB}<15$ we classified as stars all sources with $r_h<2$~pixels, consistently with \cite{Coupon2009}. 
We also cleaned the catalogue by removing all sources for which the photometry was doubtful or contaminated by neighbouring objects using a {\tt SExtractor} \citep{Bertin1996} flag greater than 3. 
We tested our selection using an eye inspection, confirming that most of the stars were removed and galaxies were in place. 

Clusters with bluer colours ($z-K \sim 2$, i.e. at redshift $\lesssim0.1$) exhibit a higher stellar contamination as many bright stars have $z-K$ colours similar to galaxies in these low redshift systems. 
We used  $class \_ star_z>0.99$ from the $z$-band catalogue provided by CFHTLS to successfully identify bright stars with blue colours ($z-K \lesssim 2$).


\section[]{Analysis}
\label{sec:analysis}

We computed the K-band luminosity for each cluster galaxy in our sample with the aim of measuring the $\rm L_K-M_{WL}$ relation. As our spectroscopic coverage is not high or uniform, we were not able to determine cluster members using a dynamical analysis \cite[see e.g.][]{Biviano2006,Ziparo2012,Ziparo2013}. Thus, we estimated a projected total K-band luminosity, i.e. via a colour selection plus statistical correction for contamination by non-members \cite[e.g.][]{Lin2003,Giodini2009}.

\subsection{K-band luminosity}
We selected candidate cluster members following the recipe of \cite{Mulroy2014}. The authors use the $(J-K)/K$ colour-magnitude space to select candidate cluster members. Near-infrared wavelengths are relatively insensitive to the star-formation and dust extinction, while (J-K) increases monotonically with redshift out to $z \sim 0.5$. This creates a narrow sequence of coeval galaxies in the $(J-K)/K$ space including both passive and star-forming sources. In the presence of a cluster, this sequence will be more populated than the field, in particular at the bright magnitudes.

\begin{figure}
\centering
  \includegraphics[trim=3cm 5cm 1cm 6cm, clip=true, angle=0, width=\hsize]{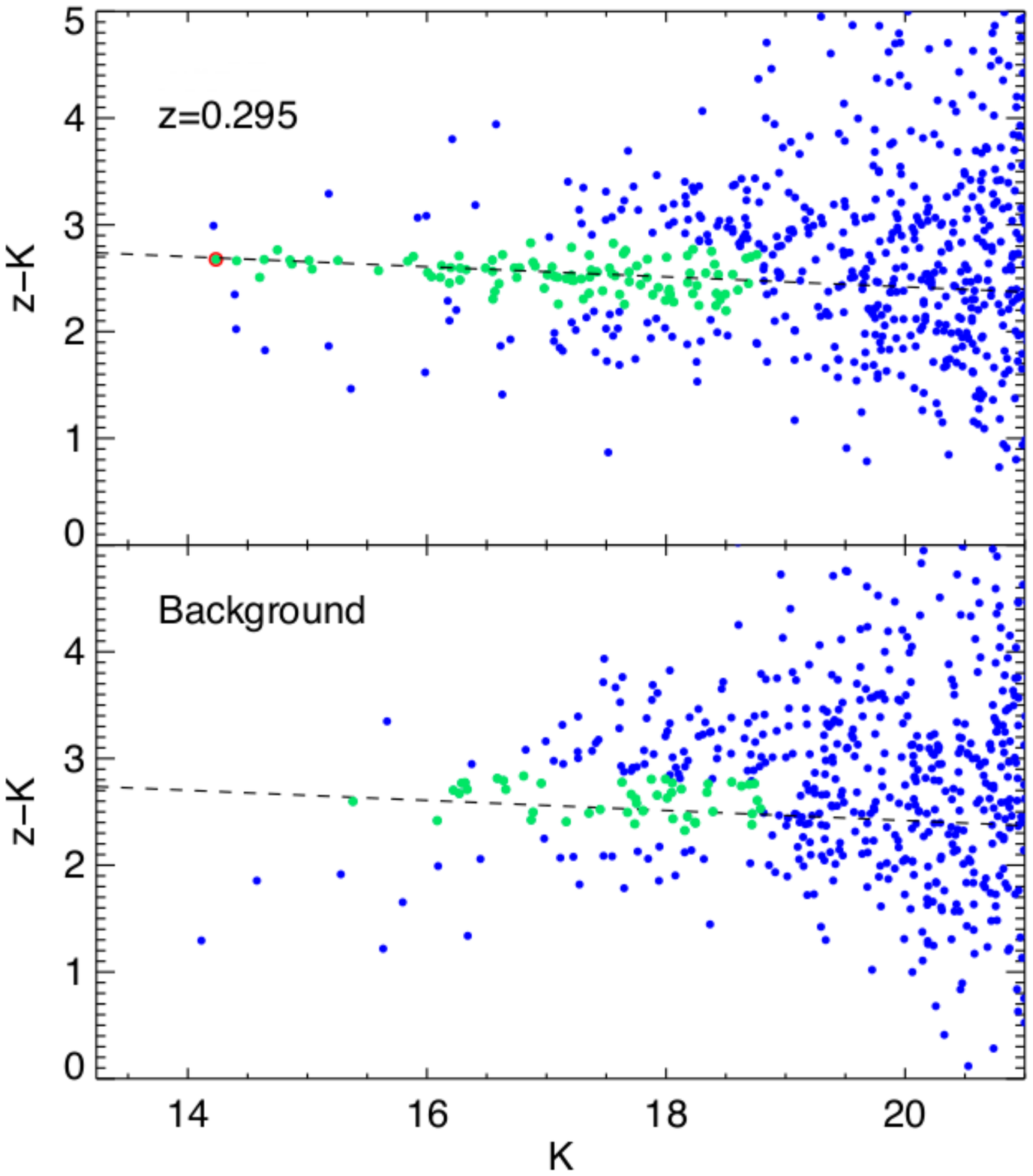}
  \caption{Colour-magnitude diagram for all galaxies lying in the projected distance of $\rm r_{500,WL}$ for XLSSC~27 (at $z=0.295$) in the XXL sample (top panel) and an area selected as background within the same projected distance (bottom panel). The dashed line shows the best linear fit for the candidate cluster members, also shown  in the bottom panel for reference. The galaxies selected to account for the total K-band luminosity of the cluster and the field are shown in green ($\rm K_{BCG}\leq K<K^*+3$ and $\rm \pm0.3~mag$ from the dashed line), while the BCG is highlighted in red. The colour sequence for the clusters appears more populated (in particular at bright magnitudes) than the background region (field) as expected.} 
  \label{fig:CMD}
\end{figure}

As J-band observations were not available for our cluster sample, we used the $(z-K)/K$ colour-magnitude diagram. We observed a similar  sequence seen by \cite{Mulroy2014} with slightly higher scatter (Fig~\ref{fig:CMD}).
For this reason, we considered all the galaxies lying within $\rm \pm0.3~mag$ (instead of $\rm \pm0.15~mag$) of the ridge line of cluster members in the $(z-K)/K$ space (Fig.~\ref{fig:CMD}) and with $\rm K_{BCG} \leq K<K^*+3$, where $\rm K_{BCG}$ is the K-band magnitude of the brightest cluster galaxy (BCG, for details on the identification see Lavoie et al., in preparation) and $\rm K^*$ is the knee of the luminosity function taken from \citealt{Lin2006}. 
Of these galaxies, we selected all sources lying within a projected distance from the X-ray centroid of $\rm r_{500,WL}$, where $\rm r_{500,WL}$ is estimated via the weak-lensing analysis, to compute $\rm L_{K,500}$. We also computed the luminosity within 1~Mpc, $\rm L_{K,1Mpc}$. 
By counting galaxies with $K<K^*+3$ we expect to miss  $\sim$10\% of the total K-band luminosity assuming a faint end slope of the cluster galaxy luminosity function $\alpha =-1.0$ \citep[e.g.][]{Mulroy2014,Balogh2011}.
We neglected the intra-cluster light (ICL) contribution in the total cluster luminosity as we expect it to contribute less than 20\% \citep{Zibetti2005}.

To convert from apparent magnitude to rest frame solar luminosity, we used the absolute K-band magnitude of the Sun $\rm K_\odot=3.39$ \citep{Kochanek2001} and the galactic extinction from the \cite{Schlegel1998} maps via the NASA/IPAC Infrared Science archive\footnote{http://irsa.ipac.caltech.edu}. Assuming that all galaxies within $\rm r_{500,WL}$ are at the redshift of the cluster, we computed a k- and evolution-correction \citep{Poggianti1997} using different simple stellar population (SSP) models from \citet[hereafter BC03]{BC03} to match our $z-K$ colours. These models assume a \cite{Chabrier2003} initial mass function (IMF) and a variety of other parameters such as the age and duration of the star burst, the metallicity of the stars, and dust extinction.
In particular, we compared three models: one with $2.5\times$ solar metallicity, one with solar metallicity, and a mixture of them from \cite{Lidman2012}. 
We note that many works use $k(r)=-6 \log(1+z)$, yielding a k$+$evolution correction very close to that predicted by the models tested in this work, but only up to $z\sim 0.3$. At $z>0.3$ the k-correction appeared too strong for $k(r)=-6 \log(1+z)$, therefore we used the BC03 models.

\begin{figure}
\centering
  \includegraphics[trim=4cm 4cm 3cm 3cm, clip=true, angle=-90, width=\hsize]{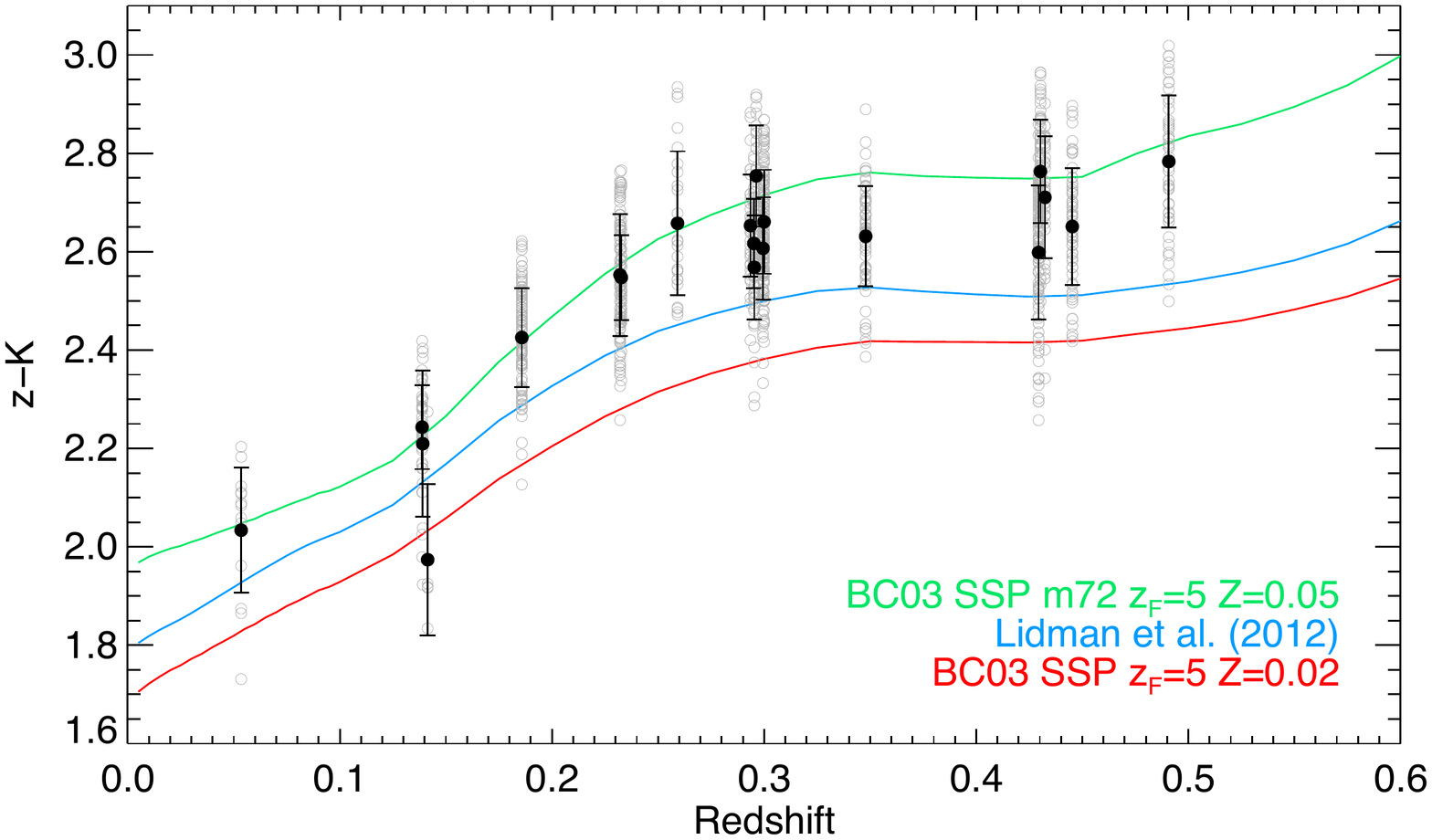}
  \caption{Observed $z-K$ colour for all candidate cluster members as a function of cluster redshift. The colours of individual galaxies are shown in grey, while the black dots and errors represent the mean colour and standard deviation for each cluster. The evolution in the $z-K$ colour for three stellar population models (see text for details) are plotted as  solid lines  as indicated in the legend.
  } 
  \label{fig:bc03_colour}
\end{figure}

The observed $z-K$ colours at different redshifts for our sample are bracketed by two models with different metallicity (Fig.~\ref{fig:bc03_colour}), one with solar metallicity and the other with Z=0.05. We derived a mean k$+$evolution correction based on these two models and we took their difference into account in the K-band luminosity error.

To remove possible interlopers we performed a statistical background correction by measuring the total K-band luminosity in 167 circular areas of the same radius used for a given system. This is the maximum number of apertures we can use that do not overlap with any cluster of our sample or with any other of the XXL-100-GC clusters. Background/field galaxies were selected with the same colour and magnitude criteria used for the clusters. The mean and standard deviation of the total $\rm L_K$ computed in the 167 areas were associated with the background luminosity and its uncertainty, respectively. We finally subtracted the background luminosity from the initial estimate of the cluster $\rm L_K$.

We derived the error on the luminosity for each cluster by adding in quadrature different components. We already mentioned the standard deviation on the background luminosities and the error of the k-correction coming from the use of different models from BC03. 
Finally, we performed a bootstrap resampling with replacement for $10^5$ resamples of the cluster candidate members before subtracting the background luminosity. The bootstrap error contribution was computed as the standard deviation of these values. 
The dominant component of the total luminosity error comes from the background subtraction, while the bootstrap resampling is the least important contribution.


\section{Results}
\label{sec:results}
In this section we present the $\rm L_K-M_{WL}$ relation for the 20 XXL clusters considered in this work. 
We measure a linear relation of the form 

\begin{equation}
         \left(\frac{L_K}{10^{12}L_\odot}\right)=\alpha+\beta \log_{10}\left(\frac{M_{WL}}{\rm 10^{14}M_\odot}\right)
,\end{equation}  

where $\alpha$ and $\beta$ are the intercept and slope, respectively. We use the publicly available IDL code of  \cite{Kelly2007} who use a Bayesian approach to linear regression. We average the asymmetric errors in mass for the fit and we measure the intrinsic scatter in the form $\rm \sigma_{ln L_K|M}$.

In this section we also perform a joint fit for the XXL and LoCuSS clusters for which masses and luminosities have been measured in a consistent way \citep{Mulroy2014}. This allows us to extend the mass range from low-mass systems to massive clusters. 

Finally, we present the richness-mass relation and the BCG light contribution to explore the galaxy population as a function of halo mass.

\subsection{K-band luminosity relation -- weak-lensing mass}
\label{sec:mass_lk}

\begin{table*}
\caption{Fit parameters for the $\rm \log_{10}(L_{K,500}/10^{12}L_\odot)=\alpha+\beta \log_{10}(M_{500,WL}/10^{14}M_\odot)$ relation.}             
\label{table:mass_lk_fit}      
\centering                          
\begin{tabular}{l c c c c c}        
\hline
\hline                 
Sample           & Radius  & $\rm n_{systems}$ & intercept         &  slope              & intrinsic scatter \\    
                     &         &                   &  $\alpha$         &  $\beta$            & $\sigma=lnL_K|M$ \\    

\hline                        
   XXL           &  $\rm r_{500,WL}$  & 20                & $0.39^{+0.17}_{-0.22}$ & $0.85^{+0.35}_{-0.27}$ & $0.37^{+0.19}_{-0.17}$        \\ 
XXL+LoCuSS       &  $\rm r_{500,WL}$  & 37                & $0.34^{+0.12}_{-0.13}$ & $1.05^{+0.16}_{-0.14}$ & $0.14^{+0.09}_{-0.07}$        \\ 
   XXL           &  1 Mpc  & 20                & $0.31^{+0.35}_{-0.43}$ & $1.00^{+0.69}_{-0.59}$ & $0.41^{+0.17}_{-0.15}$        \\ 
XXL+LoCuSS       &  1 Mpc  & 37                & $0.30^{+0.14}_{-0.16}$ & $0.97^{+0.18}_{-0.16}$ & $0.14^{+0.08}_{-0.07}$        \\ 
   XXL           &  L: 1~Mpc; M: $\rm r_{500,WL}$ & 20    & $0.56^{+0.17}_{-0.25}$ & $0.58^{+0.39}_{-0.29}$ & $0.41^{+0.16}_{-0.14}$        \\ 
XXL+LoCuSS       &  L: 1~Mpc: M: $\rm r_{500,WL}$  & 37   & $0.27^{+0.14}_{-0.17}$ & $0.99^{+0.19}_{-0.17}$ & $0.11^{+0.07}_{-0.06}$        \\ 
\hline          

\end{tabular}
\end{table*}

\begin{figure}
\centering
  \includegraphics[trim=5cm 7cm 3cm 7cm,clip=true, angle=-90,width=\hsize]{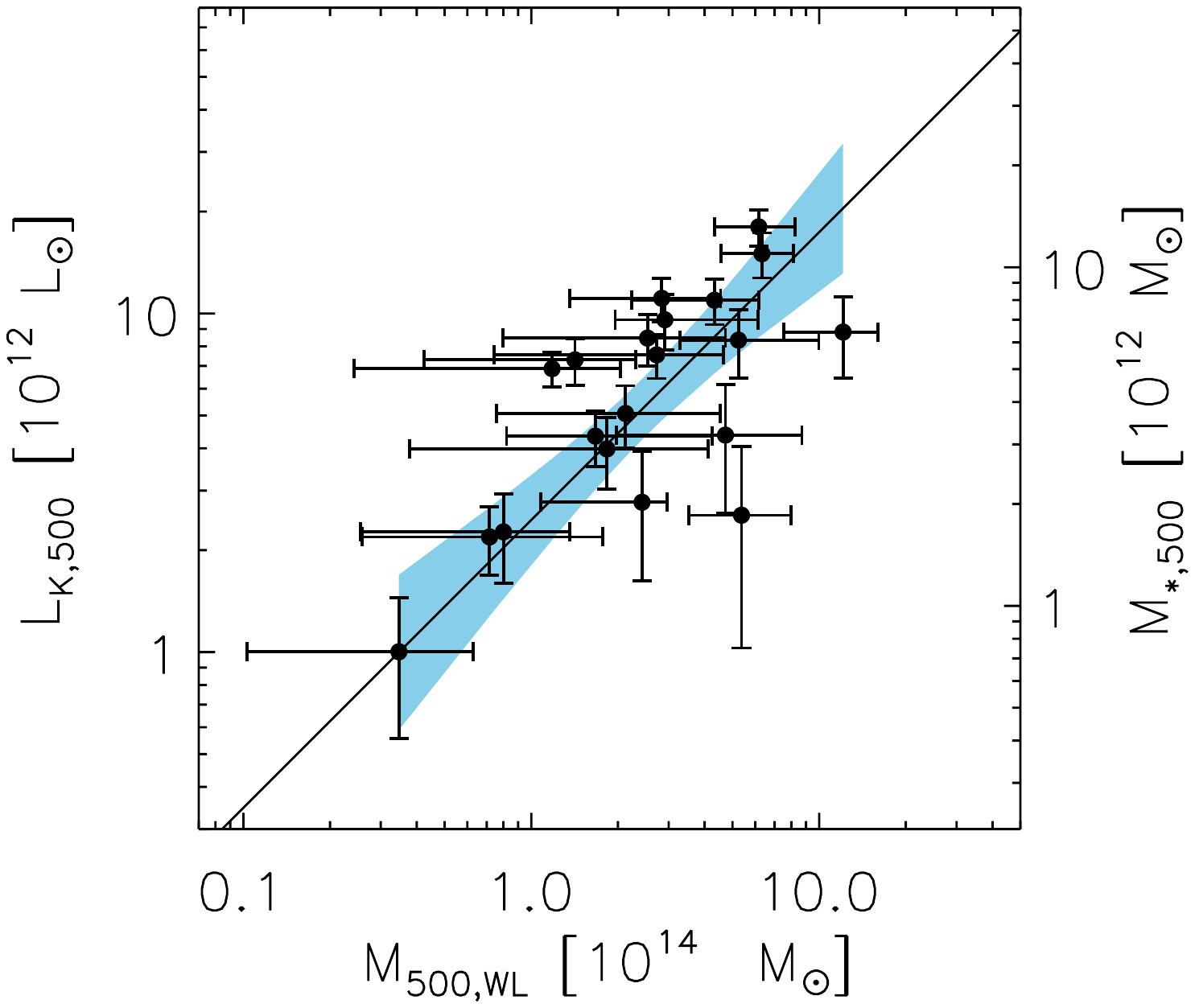}
  \caption{$\rm L_K$-mass for the 20 XXL clusters considered in this work. The black solid line and the shaded region represent the best fit and 68\% confidence interval, respectively. We use a mass-to-light ratio of 0.73 to convert $\rm L_K$ to stellar mass $\rm M_\star$. 
     }
  \label{fig:lk_mass}
\end{figure}

We use weak-lensing masses for 20 objects in our sample to compute \mlk within $\rm r_{500,WL}$ from the X-ray centroid (Fig.~\ref{fig:lk_mass} and Table~\ref{table:mass_lk_fit}). 
With this sample we extend the $\rm L_{K,500}-M_{500,WL}$ relation to the low-mass regime ($\rm M_{500,WL}=3.5\times 10^{13}\, M_\odot$) with masses calculated via weak-lensing analysis, i.e. making no assumptions on the dynamical state of the cluster.
We find that these objects are best represented by the relation 

\begin{equation}
\log_{10}\left(\frac{L_{K,500}}{10^{12}L_\odot}\right)=0.39^{+0.17}_{-0.22}+0.85^{+0.35}_{-0.27}  \log_{10}\left(\frac{M_{500,WL}}{\rm 10^{14}M_\odot}\right). 
\end{equation}To better compare our results with other works we convert $\rm L_K$ to stellar mass using a fixed mass-to-light ratio of 0.73 \citep{Cole2001} (right axis of Fig.~\ref{fig:lk_mass}).

\begin{figure*}
\centering
  \includegraphics[trim=5cm 7cm 3cm 7cm,clip=true, angle=-90,width=0.48\hsize]{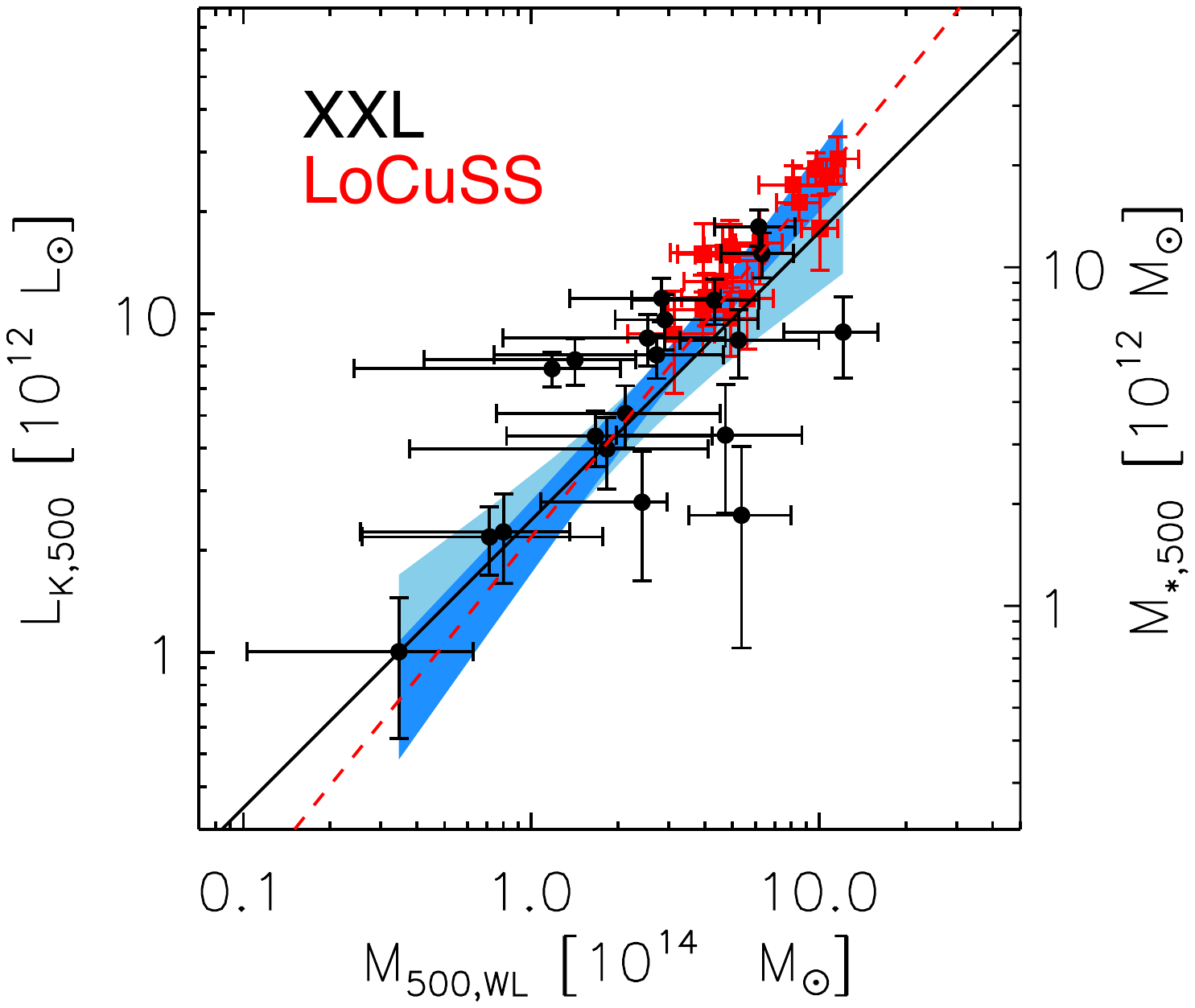}
  \includegraphics[trim=5cm 7cm 3cm 7cm,clip=true, angle=-90,width=0.48\hsize]{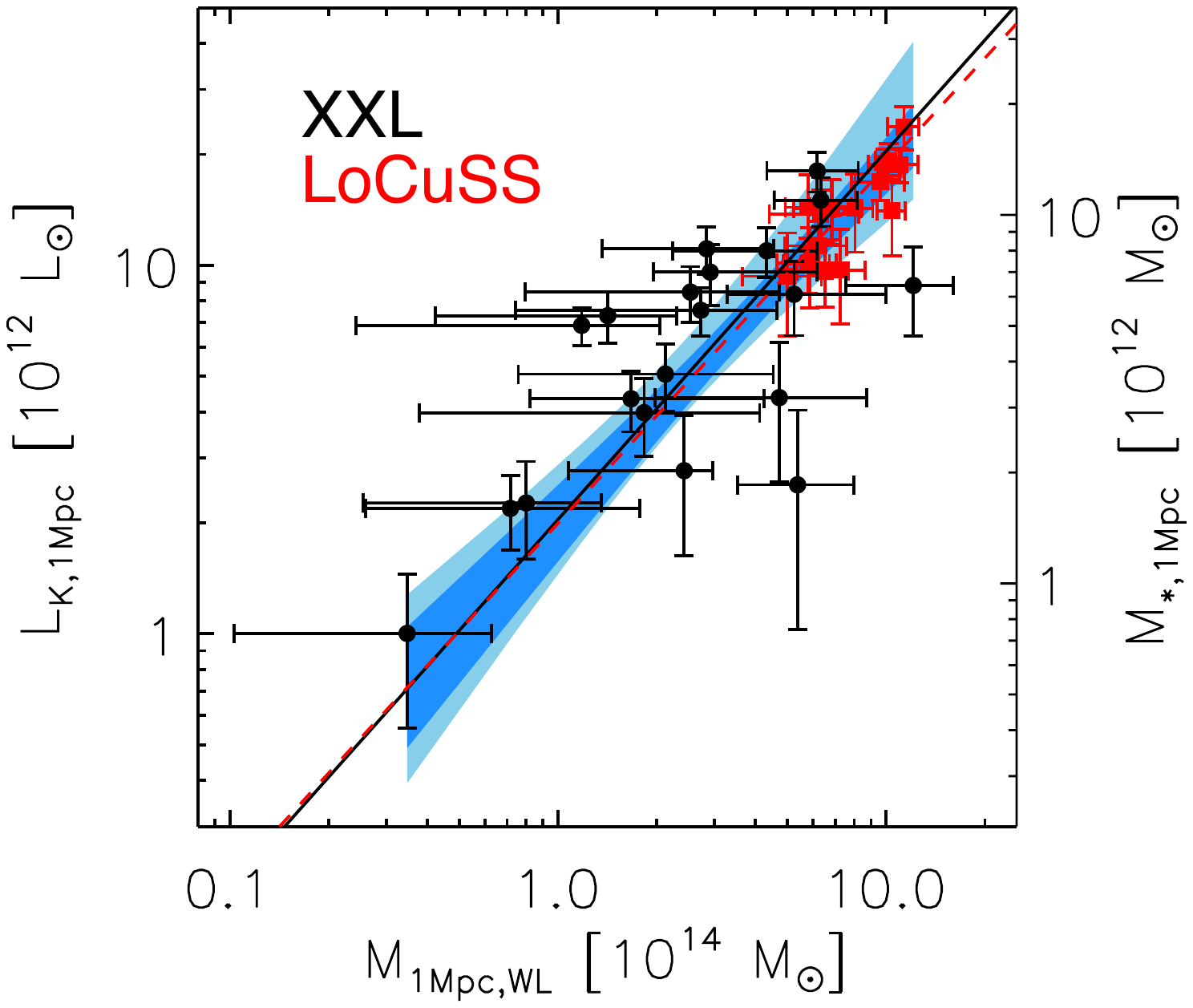}
  \caption{$\rm L_K$-mass for $\rm XXL$ (in black) and LoCuSS (in red) clusters (from \citealt{Mulroy2014}) using an aperture of $\rm r_{500,WL}$ (left panel) and 1~Mpc (right panel). The black solid line and the light blue shaded region are the best fit and 68\% confidence interval for XXL only points, while the red dashed line and the darker blue shaded region show the joint XXL and LoCuSS fit and the corresponding 68\% confidence interval. 
   }
  \label{fig:lk_mass_locuss}
\end{figure*}

To populate the massive end of the \mlk and increase the statistics, we consider other results on \mlk based on weak-lensing analysis. 
As our analysis and the work on massive clusters in the LoCuSS sample \citep{Mulroy2014} are performed in a similar way, we jointly fit the $\rm L_K-M_{WL}$ relation for our clusters and the systems in LoCuSS. We multiply LoCuSS masses by 1.2 to account for the 20\% bias in the mass measurement found by \cite{Okabe2013}. As the luminosities are measured within $\rm r_{500,WL}$ (i.e. there is  dependence on the mass) we also multiply the LoCuSS luminosities by $\rm 1.2^{1/b}$, where b is the best fit slope measured by \cite{Mulroy2014} for this set of points. 
The joint fit yields slope and intercept consistent with values estimated for the XXL-only sample (left panel of Fig.~\ref{fig:lk_mass_locuss}):

 \begin{equation}
 \log_{10}\left(\frac{L_{K,500}}{10^{12}L_\odot}\right)=0.34^{+0.12}_{-0.13}+1.05^{+0.16}_{-0.14} \log_{10}\left(\frac{M_{500,WL}}{\rm 10^{14}M_\odot}\right).
 \end{equation}The inclusion of the LoCuSS points in the fit allows us to measure a lower intrinsic scatter $\rm \sigma_{ln L_K|M}=0.14^{+0.09} _{-0.07}$ and lower parameter uncertainties with respect to the XXL-only fit (Table~\ref{table:mass_lk_fit}).

Measuring $\rm L_K$ and $\rm M_{WL}$ within $\rm r_{500,WL}$ introduces an intrinsic correlation between these quantities because $\rm r_{500,WL}$ scales with $\rm M_{500}^{1/3}$.  To check the impact of this intrinsic correlation on our results we first measure $\rm L_K$ within a fixed physical aperture of radius 1~Mpc, following \cite{Mulroy2014} and \cite{Lin2004}.
The fit between $\rm L_{K,1Mpc}$ and $\rm M_{500,WL}$ for the XXL sample gives a slope of $0.58^{+0.39} _{-0.29}$, which is much shallower than the previous fits (Table~\ref{table:mass_lk_fit}). Conversely, when fitting XXL and LoCuSS points together, we find a slope consistent with unity, i.e. much different from the value we obtain for XXL only.
The main difference in the two samples, is that XXL spans a wide mass range (with $\rm r_{500,WL}=0.5-1.4 Mpc$), while LoCuSS has $\rm r_{500,WL}$ of the order of 1~Mpc.
Thus, luminosity and mass are measured within very different regions for the XXL sample. In other words, a fixed aperture is not ideal for measuring the luminosity of clusters spanning a wide size range.

A like-for-like comparison is given by $\rm M_{WL}$ and $\rm L_K$ measured within the same fixed aperture of $\rm 1~Mpc$ (right panel of Fig.~\ref{fig:lk_mass_locuss}). Although the uncertainty on the fit parameters increases for the XXL-only sample, we find a slope of unity perfectly consistent with the joint XXL+LoCuSS fit.

\subsection{Richness and BCG contribution}

\begin{figure}
\centering
  \includegraphics[trim=5cm 6.5cm 3cm 9cm,clip=true, angle=-90, width=\hsize]{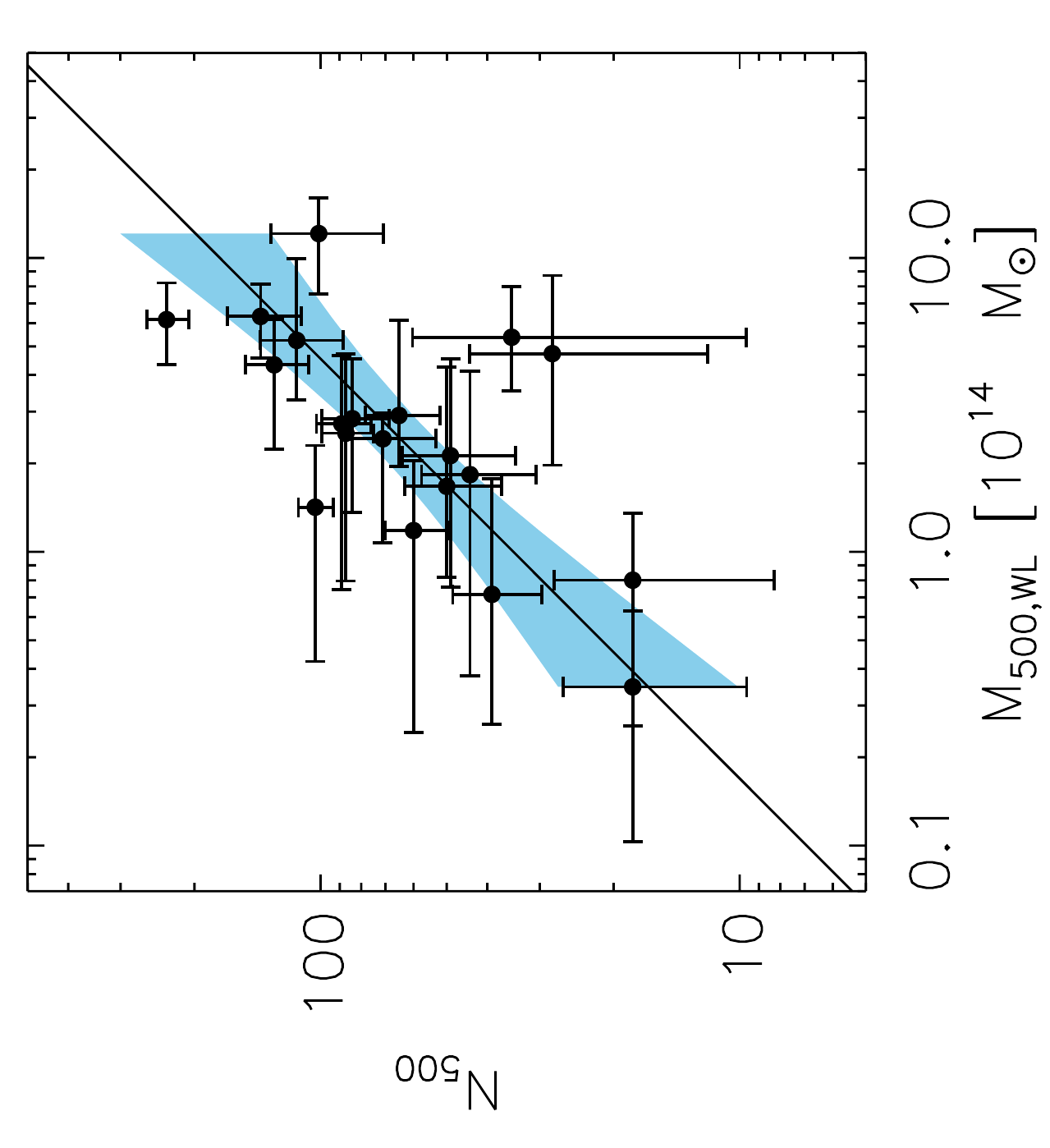}
  \caption{Number of galaxies as a function of weak-lensing mass for the $\rm XXL$ sample considered in this work. The black solid line shows the best fit relation represented by $\rm \log_{10} N_{500}=1.54^{+0.13} _{-0.19} +  0.70^{+0.29} _{-0.23}\, \log_{10}(M_{500,WL}/10^{14}M_\odot$). The shaded region shows the 68\% confidence interval. 
 }

  \label{fig:hod}
\end{figure}

The joint XXL+LoCuSS fit is consistent with star formation efficiency being independent of halo mass across the range of cluster and group masses probed by our sample.
However, as we cannot exclude a slope shallower than unity at $>2\sigma$, we investigate this issue from another point of view. We study the richness--weak-lensing mass ($\rm N_{500}-M_{500,WL}$) relation (Fig.~\ref{fig:hod}) for the XXL systems, where $\rm N_{500}$ is the number of all galaxies within $\rm r_{500,WL}$ that contribute to the cluster $\rm L_{K,500}$. 
The XXL systems follow a positive correlation with a best fit of
\begin{equation}
        \log_{10} N_{500}=1.54^{+0.13} _{-0.19} +  0.70^{+0.29} _{-0.23}\, \log_{10}\left(\frac{M_{500,WL}}{\rm 10^{14}M_\odot}\right). 
\end{equation}

The slope is slightly shallower than that found by \cite{Lin2004}, but consistent within the errors.
We measure an intrinsic scatter of $\rm \sigma_{lnN|M}=0.33^{+0.16} _{-0.14}$, lower than that measured by \cite{Lin2004}.

\begin{figure*}
\centering
  \includegraphics[trim=5cm 7cm 3cm 9cm,clip=true, angle=-90, width=0.48\hsize]{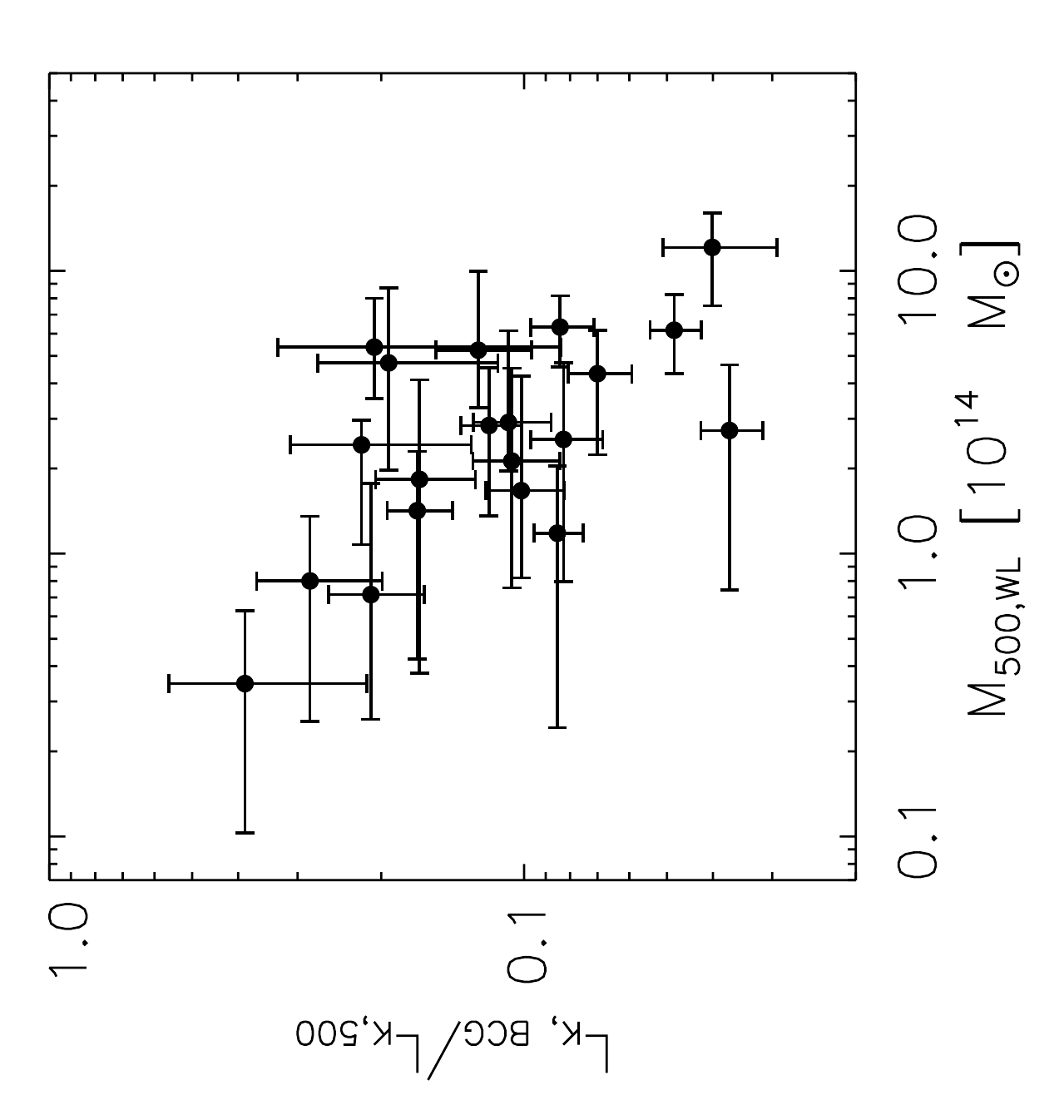}
  \includegraphics[trim=5cm 7cm 3cm 9cm,clip=true, angle=-90, width=0.48\hsize]{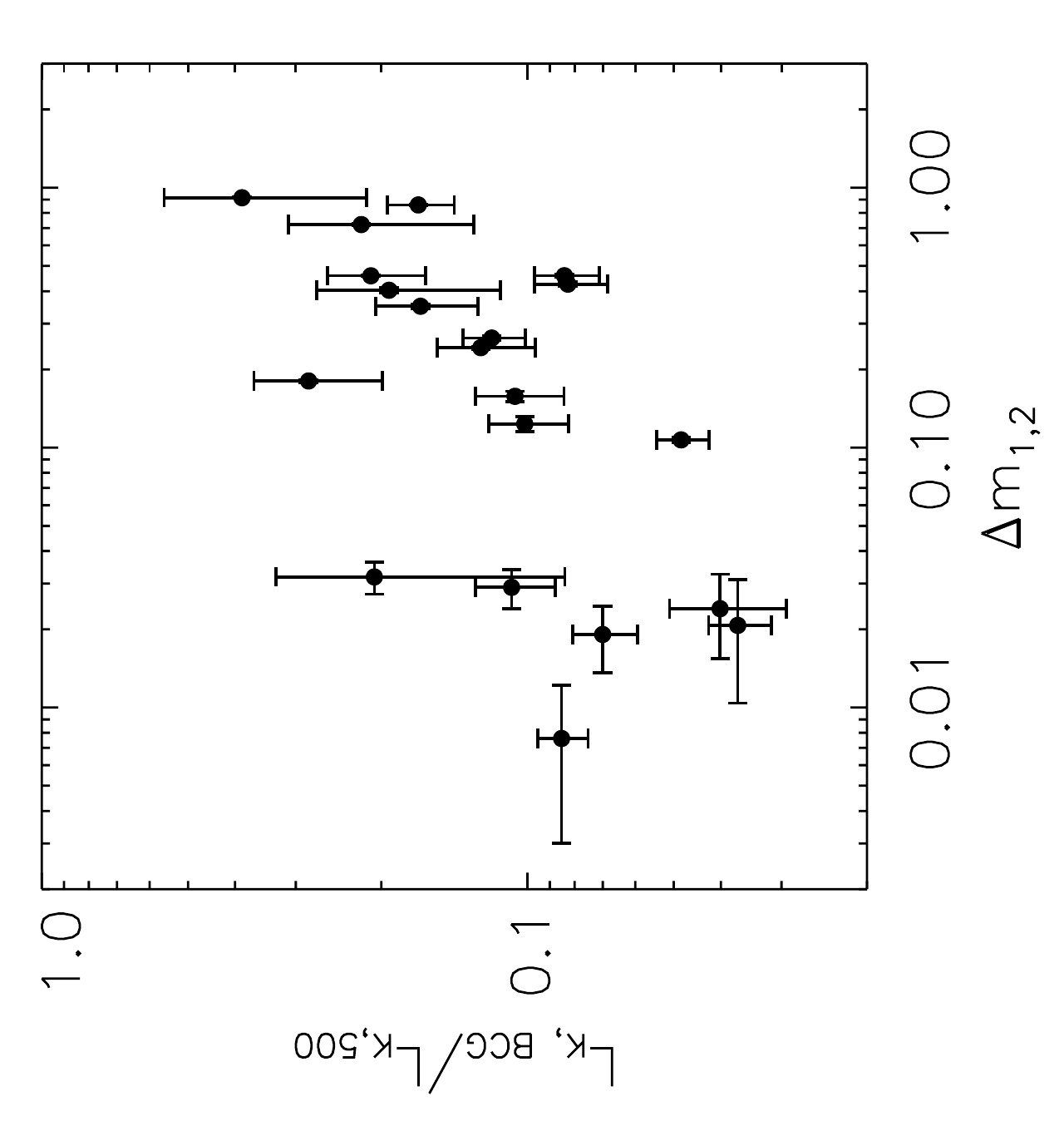}
  
  \caption{Fractional contribution in luminosity of the BCG as a function of weak-lensing mass (left panel) and as a function of the magnitude gap between the two brightest members of the clusters (right panel). 
}
  \label{fig:mass_lk_bcg}
\end{figure*}

While the K-band luminosity scales with the halo mass with a slope consistent with unity, the $\rm N_{500}-M_{500,WL}$ relation exhibits a shallower slope. This means that the cluster K-band luminosity steadily increases with the halo mass, whereas less massive clusters are populated, on average, by a larger number of galaxies than expected for more massive clusters. This implies that group-sized halos host less luminous/massive galaxies than their massive counterparts. 

We investigate the relative contribution of the BCG to the K-band luminosity of the cluster (left panel of Fig.~\ref{fig:mass_lk_bcg}). The fractional BCG contribution and the halo mass are anti-correlated with a Spearman correlation coefficient of -0.52 and a probability that this happens by chance of $1.18 \times 10^{-2}$. This suggests that the BCG fractional light contribution is more important for low-mass clusters ($\rm L_{K,BCG} /L_{K,500} \sim 0.4$) than for massive ones ($\rm L_{K,BCG} /L_{K,500} \sim 0.04$), in agreement with \cite{Lin2004BCG} and \cite{Gonzalez2013}.

The higher BCG contribution to the stellar budget of low-mass clusters can be reconciled with the number of galaxies being higher, in particular if galaxies in low-mass systems are less luminous than those in massive ones. In fact, one would expect  galaxies to grow in stellar mass via merging and star formation, decreasing the number of galaxies. 
To investigate this, we explore the fractional light contribution of the BCG as a function of the magnitude gap between the BCG and the second brightest member (right panel of Fig.~\ref{fig:mass_lk_bcg}). We select the brightest galaxies within the colour selection we use for the cluster candidate members before performing the background subtraction. 
The Spearman test reveals a coefficient of 0.60 with a probability of non-correlation of $4.37\times 10^{-3}$, i.e. the light contribution of the BCG increases with the luminosity gap. 
This can be expected if we consider the effects of dynamical friction: a galaxy of given mass is dragged on a shorter timescale to the centre of a group than  a more massive cluster is. Thus, dynamical friction easily explains the larger magnitude gap (right panel of of  Fig.~\ref{fig:mass_lk_bcg}) and the higher contribution of the BCG to the system light (left panel of Fig.~\ref{fig:mass_lk_bcg}) in groups rather than in clusters.

\section{Discussion}
\label{sec:discussion}

We have investigated the \mlk relation for 20 galaxies of the XXL-100-GC sample with a wide range of masses and redshifts. For the first time, we have measured this relation using weak-lensing analysis to estimate cluster masses down to $\rm M_{500,WL} =3.5 \times 10^{13}\,M_\odot$. This has allowed us to investigate whether the slope of the $\rm L_K-M$  relation is a function of halo mass or whether previous claims of different star formation efficiencies for group-sized halos compared to massive clusters are due to a bias in the mass measurements.
We find a positive correlation between K-band luminosity and mass with a slope of $0.86^{+0.37}_{-0.28}$ and an intrinsic scatter $\rm \sigma_{ln L_K|M} = 0.37^{+0.19}_{-0.17}$. 
With the aim of increasing the statistics and populating the massive end of the \mlk relation, we have increased our sample with clusters from the LoCuSS survey presented by \cite{Mulroy2014}.  As these authors use weak-lensing masses and compute near-infrared luminosities in a similar way to our own method, we have performed a joint fit, confirming a slope consistent with unity ($1.05^{+0.16}_{-0.14}$) and obtaining an intrinsic scatter $\rm \sigma_{ln L_K|M} = 0.14^{+0.09}_{-0.07}$. 

\begin{figure}
\centering
  \includegraphics[trim=5cm 7cm 3cm 7cm,clip=true, angle=-90, width=\hsize]{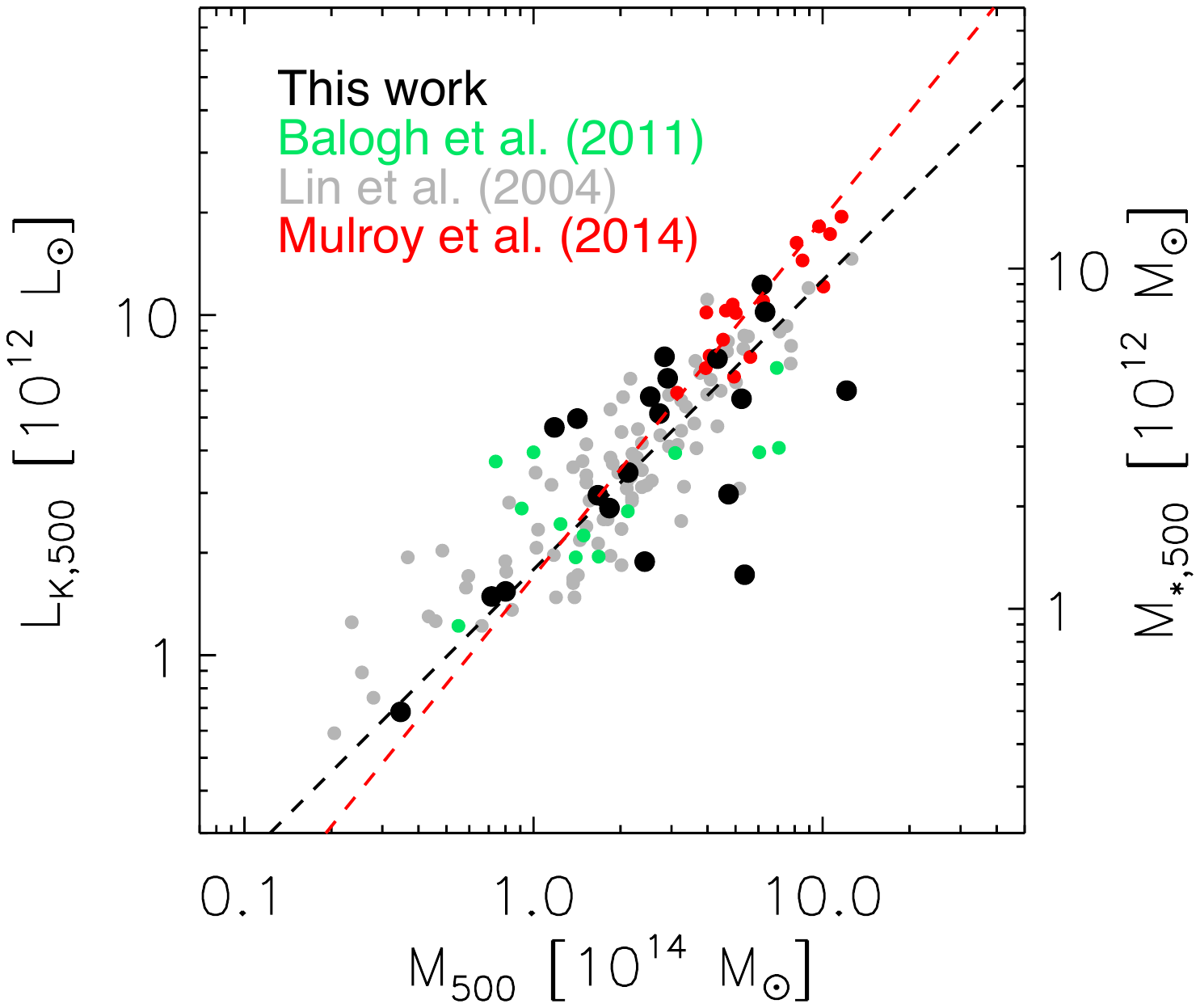}
  \caption{$\rm M_{500}-L_{K,500}$  relation for this work and other works in the literature as indicated by the legend. The dashed lines show the best linear fit for XXL only (black) and XXL+LoCuSS (red). \citet{Balogh2011} and \citet{Lin2004} measure 3D luminosities, while we de-project $\rm L_K$ for XXL and LoCuSS \cite[as in][]{Mulroy2014} by multiplying these luminosities by 0.68 (see text for details).
}
  \label{fig:mass_lk_comp}
\end{figure}

We compare our results with previous work in the literature with cluster masses and luminosities computed in different ways (Fig.~\ref{fig:mass_lk_comp}). \cite{Lin2004} derive masses using the mass-temperature relation and integrate the luminosity function down to $\rm K^*+3$ to derive the total $\rm L_K$. \cite{Balogh2011} perform a dynamical analysis to retrieve system masses and cluster members; they sum up individual member luminosities to compute the cluster near-infrared luminosities. 
For the sake of comparison, we de-project our total K-band luminosities i.e. we convert the luminosities measured within a cylinder to luminosities on a spherical space. We follow \cite{Mulroy2014} who multiply their $\rm L_K$ by 0.68, a value estimated by the weak-lensing analysis of \cite{Okabe2010}. After de-projecting the $\rm L_K$ we find a general agreement with the other authors (Fig.~\ref{fig:mass_lk_comp}).

The $\rm L_{K,500}-M_{500}$ seems to flatten for group-sized halos, in particular for systems for which the mass has not been computed via weak-lensing analysis. \cite{Lin2004} find a slope for the $\rm L_{K,500}-M_{500}$ of $0.69\pm 0.04$, arguing that low-mass systems are more efficient at forming stars than their massive counterparts. 
Although we cannot completely rule out this scenario, the slope of the \mlk we measure is also consistent with unity, suggesting that the star formation efficiency might be independent of the mass of the system. 
The direct comparison with the findings of \citet[][see Fig.~\ref{fig:mass_lk_comp}]{Lin2004} suggests that the flattening of the $\rm L_{K,500}-M_{500}$ at low masses might be due to a bias in the mass measurement. While the weak-lensing masses used in this work do not rely on any assumptions about the dynamical state of the clusters, \cite{Lin2004} assume hydrostatic equilibrium as they derive cluster masses from an X-ray scaling relation. 

Several authors in the literature have used stellar mass to trace halo mass. We have used a fixed mass-to-light ratio to convert $\rm L_K$ into stellar mass. Our XXL+LoCuSS slope is at odds with \cite{Giodini2009} who find a slope of $0.72 \pm 0.13$ for a secure sample of low-mass X-ray groups in the COSMOS field (although our result is marginally consistent with their slope of $0.81 \pm 0.11$ for their entire sample).  \cite{Patel2015} complement the sample of \cite{Giodini2009} with 23 low-mass X-ray groups, finding consistent results with \cite{Giodini2009}, but with a larger scatter (0.25~dex). For both works the stellar masses come from a multiwavelength template fit while the halo masses are estimated using the mass-X-ray luminosity scaling relation of \cite{Leauthaud2010}. 
Recent results \citep{Balogh2014, VanDerBurg2014} based on systems at $z\sim 1$ with halo masses derived via dynamical analysis show consistency with \cite{Giodini2009} and \cite{Patel2015}. 

Our results are also at odds with \cite{Kravtsov2014} who analyse a set of nine clusters. These authors find a close relation (scatter of $\rm \sim 0.1~dex$) between stellar mass and halo mass and a slope of $0.59 \pm 0.08$ . 
We note that their sample is complemented with that of \cite{Gonzalez2013}, thus the total sample avoids highly disturbed systems, and estimates halo masses from X-ray temperature or by using the $\rm Y_X$ parameter. These results are much closer to those of \cite{Andreon2010} than ours. Indeed, \cite{Andreon2010} find a slope of $0.45 \pm 0.08$ for a sample of local clusters for which dynamical masses are available. We note that  of the aforementioned works, the only one that  relies on weak-lensing masses is \cite{Mulroy2014}.

Recently, \cite{Hilton2013} have studied the stellar-to-halo mass relation for a sample of 14 clusters. Using observations at 3.6~$\mu$m and 4.5~$\mu$m and the  background subtraction technique, they find a slope of 0.9$\pm$0.4, which is consistent with our results.

To avoid any dependence on the cluster mass, we have also measured the K-band luminosity for our systems within a fixed aperture of 1~Mpc. We have followed the example of \cite{Mulroy2014} and \cite{Lin2004} who discuss the potential use of the galaxy K-band light as a proxy for cluster mass. The different slopes found for $\rm L_{K,1Mpc}-M_{500,WL}$ using XXL only and XXL+LoCuSS samples suggest that measuring the luminosity in a fixed aperture for systems with a wide mass range is not ideal. While $\rm r_{500,WL}$ for low-mass XXL systems is of the order of $\sim 500$~kpc, massive clusters have $\rm r_{500,WL}\gtrsim 1$~Mpc. Thus, $\rm M_{WL}$ and $\rm L_K$ are measured in different region sizes. This becomes clearer when we measure the $\rm L_{K,1Mpc}-M_{1Mpc,WL}$ within a fixed aperture of 1~Mpc for both masses and luminosities. We find a slope of unity, consistent with \cite{Mulroy2014}.

The issues mentioned above do not apply to \cite{Mulroy2014} who find similar slope and slightly higher scatter in their $\rm L_{K,1Mpc}-M_{500,WL}$ relation compared to their $\rm L_{K,500}-M_{500,WL}$. In fact, their sample spans a narrower range of masses than ours with typical cluster radii being of the order of 1~Mpc.

\subsection{Richness and BCG contribution}

We have investigated the richness as a function of weak-lensing mass for our sample finding a slope shallower than that of the \mlk relation. The slope we find ($0.70^{+0.29} _{-0.23}$) is slightly shallower than, but not inconsistent with, the value reported by \cite{Lin2004} who measure a slope of $0.82 \pm 0.04$. These results indicate that low-mass systems host more galaxies than expected compared to more massive systems.

Our findings are in full agreement with the observed $\rm N_{500}-M_{500}$ relation of \cite{Poggianti2010} who find a slope of $0.77 \pm 0.03$. \cite{Poggianti2010} also find that this slope is dominated by systematic errors in their mass measurements. In fact, while they expect a slope of $1.00 \pm 0.04$ from their simulated clusters, they find consistent results with observations when they introduce selection effects in their simulated data. The authors argue that projection effects may bias their observational estimate of the halo masses. Although we cannot exclude  projection effects (e.g. clusters elongated along the line of sight), our slope of $0.70^{+0.29} _{-0.23}$ is marginally consistent, within the errors, with unity. 

The different slopes of \mlk and $\rm N_{500}-M_{500,WL}$ suggest that, while the total near-infrared luminosity increases steadily with halo mass, some processes act to decrease (or not increase) the number of galaxies. Good candidates are galaxy merging, i.e. the galaxies increase their stellar mass, thus their $\rm L_K$, and decrease in number. 
The most obvious objects to be formed via merging are the BCGs \cite[e.g.][]{Dubinski1998}. 
More specifically, the process known  as {\it galactic cannibalism} \citep{White1976, Ostriker1977}  indicates that BCGs mainly increase their mass via dissipationless merging events of galaxies already in place at higher redshift.
According to \cite{Lidman2012} the build-up of stellar mass in BCGs mainly occurs through major mergers with a stellar mass increase of a factor of $\sim 1.5$ over $0<z<1$ \citep[a factor of $\sim3$ according to][]{Laporte2013}. 
This scenario is supported by \cite{Behroozi2013} who study the stellar-to-halo mass relation using the abundance matching technique. They find that galaxies more massive than the Milky Way grow via mergers at $0<z<1$, while galaxies less massive increase their mass via star formation activity.

We have found that the relative light contribution of the BCG to the K-band luminosity of the cluster (left panel of Fig.~\ref{fig:mass_lk_bcg}) is much more important for group-sized halos than for massive clusters, in agreement with \cite{Lin2004BCG}.  This has also been noted by \cite{Ziparo2013} who find mass segregation only in the inner regions of galaxy groups, most likely due to the dominance of the BCG.
To reconcile the slopes of \mlk and $\rm N_{500}-M_{500,WL}$, bright galaxy members (excluding the BCG) should be, on average, more luminous in massive clusters than in groups.  
This scenario is confirmed by the correlation between the fractional light contribution of the BCG with the luminosity gap (right panel of Fig.~\ref{fig:mass_lk_bcg}). Indeed, the magnitude gap shows the difference in luminosity between the first two brightest galaxies in a cluster, implying that its amplitude is a function of both the formation epoch and the recent infall history of the cluster \citep{Smith2010}.

According to the pre-processing \citep{Zabludoff1998} mechanism, massive clusters grow via the accretion of smaller systems where galaxies gain both light and mass \citep[e.g.][]{McGee2009,Berrier2009}. Observational evidence has been found from the identification of groups embedded in the large-scale structure of clusters \citep{Cortese2006, Tanaka2007, Ziparo2012, Eckert2014}.
This would explain, as already suggested by \cite{Lin2004BCG}, the decreasing light contribution of the central brightest galaxies in clusters: BCGs would form in low-mass systems and then become bright galaxies in clusters as part of the hierarchical structure formation \citep{Merritt1985,Edge1991,DeLucia2007,DeLucia2012}.

\section{Summary and conclusions}
\label{sec:conclusions}

We have explored the $\rm L_K-M_{WL} $ ($\rm M_\star-M_{WL}$) relation for a subsample of the the XXL-100-GC observed with WIRCam. For the first time, we have measured this relation using weak-lensing analysis to estimate cluster masses down to $\rm M_{500,WL} =3.5 \times 10^{13}\,M_\odot$. We summarise our results below.

\begin{itemize}
\item \mlk has a slope of $0.86^{+0.37}_{-0.28}$, consistent with unity, and an intrinsic scatter of $\rm \sigma_{lnM|L_K}=0.37$. Extending our study to massive clusters from LoCuSS \citep{Mulroy2014} yields a slope of $1.05^{+0.16}_{-0.14}$ and a scatter of $\rm \sigma_{lnM|L_K}=0.14$.

\item Comparisons with previous studies based on masses estimated via X-ray properties or derived from scaling relations, in particular for low-mass systems, suggest that star formation efficiency is independent of halo mass. 

\item The different slopes obtained for $\rm L_{K,1Mpc}-M_{500,WL}$ and $\rm L_{K,1Mpc}-M_{1Mpc,WL}$ show that measuring the luminosities within a fixed aperture is not ideal for samples with a wide mass range like XXL. In fact, XXL includes a variety of cluster sizes with $\rm r_{500,WL}$ being much different from the radius within which the luminosity is measured.

\item The $\rm N_{500}-M_{500,WL}$ relation yields a slope shallower than unity. This suggests that group-sized halos host a higher number of galaxies compared to that expected from their massive counterparts. To reconcile the \mlk and the $\rm N_{500}-M_{500,WL}$ we conclude that galaxies in groups are less luminous than cluster galaxies. 

\item BCGs contribute more to the total luminosity budget of low-mass systems than their massive counterparts. The dominance of the BCG increases with the luminosity gap, suggesting that BCGs form and evolve in groups before falling into massive clusters where they become bright galaxies or BCGs in clusters as part of the hierarchical structure formation.

\end{itemize}

This work is based on 20 clusters in the XXL Survey. In order to have more robust estimates of the quantities explored so far, we need to increase our sample. In a future work, we plan to include clusters observed by the VIDEO\footnote{http://star-www.herts.ac.uk/~mjarvis/video/index.html} and UKIDSS\footnote{http://www.ukidss.org} surveys to improve statistical precision, using all weak-lensing masses presented in \cite{Lieu2015}.
Moreover, we will expand our sample of clusters with accurate weak-lensing masses as more weak-lensing data becomes available for both the northern and southern fields in XXL. 
Finally, the upcoming analysis of 50 clusters from LoCuSS (Mulroy et al., in prep.) will allow us to extend the sample to high masses, yielding a more robust estimate of the fit parameters of the \mlk relation.

\begin{acknowledgements}
        
We thank the anonymous referee for the constructive comments.

        XXL is an international project based around an XMM Very Large Programme surveying two $\rm 25~deg^2$ extragalactic fields at a depth of $\rm \sim 5 \times 10-15\, erg\, cm^{-2} s^{-1}$ in the [0.5-2] keV band for point-like sources. The XXL website is http://irfu.cea.fr/xxl. Multi-band information and spectroscopic follow-up of the X-ray sources are obtained through a number of survey programmes, summarised at http://xxlmultiwave.pbworks.com/.
        
We thank the CFHT staff and all the PIs that have contributed to the success of this work (K. Thanjavur, G. Morrison, L. VanWaerbeke,  J. P. Willis and S. Arnouts).

        Based on observations obtained with MegaPrime/MegaCam, a joint project of CFHT and CEA/DAPNIA, at the Canada-France-Hawaii Telescope (CFHT) which is operated by the National Research Council (NRC) of Canada, the Institut National des Sciences de l'Univers of the Centre National de la Recherche Scientifique (CNRS) of France, and the University of Hawaii. This work is based in part on data products produced at TERAPIX and the Canadian Astronomy Data Centre as part of the Canada-France-Hawaii Telescope Legacy Survey, a collaborative project of NRC and CNRS.

        Based on observations obtained with WIRCam, a joint project of CFHT, Taiwan, Korea, Canada, France, and the Canada-France-Hawaii Telescope (CFHT) which is operated by the National Research Council (NRC) of Canada, the Institute National des Sciences de l'Univers of the Centre National de la Recherche Scientifique of France, and the University of Hawaii.

This work is based (in part) on data products produced at the TERAPIX data center located at the Institut d'Astrophysique de Paris.

FZ and GPS acknowledge support from the Science and Technology Facilities Council.
FP acknowledges support from the BMBF/DLR grant 50 OR 1117, the DFG grant RE 1462-6 and the DFG Transregio Programme TR33.

FZ thanks Chris Haines and Nobuhiro Okabe for useful discussions.
\end{acknowledgements}


\end{document}